# Design and experimental validation of an active catheter for endovascular navigation

**Thibault Couture**
Service de Chirurgie vasculaire, Hôpital Pitié-Salpêtrière
52 Boulevard Vincent-Auriol, 75013 Paris
thibault.couture@gmail.com

**Jérôme Szewczyk**[*]
Institut des Systèmes Intelligents et de Robotique, Université Pierre et Marie Curie
Boîte courrier 173, 4 place Jussieu 75252 Paris Cedex 05, France
szewczyk@isir.upmc.fr

**ABSTRACT**

*Endovascular technique has many advantages but relies strongly on operator skills and experience. Robotically steerable catheters have been developed but few are clinically available. We describe here the development of an active and efficient catheter based on Shape Memory Alloys (SMA) actuators.*

*We first establish the specifications of our device considering anatomical constraints. We then present a new method for building active SMA-based catheters. The proposed method relies on the use of a core body made of three parallel metallic beams and integrates wire-shaped SMA actuators. The complete device is encapsulated into a standard 6F catheter for safety purposes. A trial-error campaign comparing 70 different prototypes is then conducted to determine the best dimensions of the core structure and of the SMA actuators with respect to the imposed specifications. The final prototype is tested on a silicon-based arterial model and on a 23-kg pig. During these experiments we were able to cannulate the supra-aortic trunks and the renal arteries with different angulations and without any complication.*

---

[*] Corresponding author

*A second major contribution of this paper is the derivation of a reliable mathematical model for predicting the bending angle of our active catheters. We first use this model to state some general qualitative rules useful for an iterative dimensional optimization. We then perform a quantitative comparison between the actual and the predicted bending angles for a set of 13 different prototypes. It happens that the relative error is less than 20% for bending angles between 100° and 150° which is the interval of interest for our applications.*

**INTRODUCTION**

Cardiovascular diseases are the first cause of death in the world according to the World Health Organization. The most frequent pathologies are myocardial infarction, stroke, and peripheral vascular diseases. The treatment of the underlying lesions (stenosis of coronary, carotid and other peripheral vessels) can be performed using surgical or endovascular techniques. The last option has rapidly developed over the last two decades. It relies on the use of catheters and guide wires to reach pathological lesions and deliver stents or balloons to treat them. Endovascular techniques are less invasive approaches than conventional surgeries, allowing early postoperative recovering, a decreased risk of complications and reduced operating times. Despite all these advantages, the success of a procedure often relies on the operator skills and experience. Patient challenging anatomy is the most frequent cause of difficulties or failure. Approximately one in eight patients have a renal artery take-off angle that is less than 50° [1]. In these patients, an approach via a femoral access site can be technically challenging and may result in an unsuccessful procedure. In most cases, several material exchanges are needed in one procedure, increasing the operative time. Several endovascular robotic systems have been developed to improve catheter navigation and reduce both the operating time and material.

**Extracorporeal robots**

Extracorporeal robots constitute a first kind of endovascular robotic systems (Fig. 1). This category includes robotically steerable catheters like the Magellan robot (Hansen Medical, Mountain View, Calif) and magnetically controlled systems like the Niobe system (Stereotaxis, St. Louis, Mo).

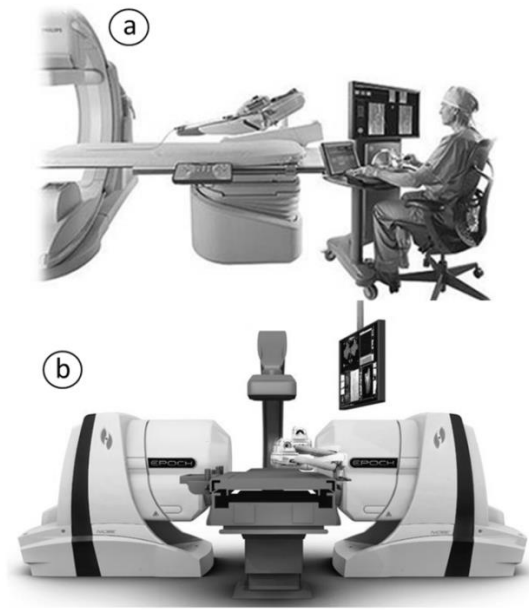

Fig. 1. Extracorporeal robots: a- Magellan®, b- Niobe® systems

The main advantages of extracorporeal systems are: reduced number of catheter exchanges, enhanced catheter stability and positional control, reduced contrast and fluoroscopy time, reduced requirements in manual skill and shorter learning curve [2-3]. The Magellan system has been studied on animal models and through clinical studies [4–7]. Its advantages in terms of catheterization times, absolute number of movements of the catheter tip and number of wall hits have been demonstrated in-vitro [8–10]. But in a recent literature review on endovascular navigation systems [11], the authors concluded that there were not enough proof of the superiority of such robotically driven catheters over classical techniques and that robotic systems should be reserved only in case of conventional techniques failures.

A meta-analysis [12] in 2013 on magnetically controlled systems reports decreased scopy times, a superior safety and efficiency during cardiac rhythm trouble ablation compared to conventional catheters. However, high cost, high volume, long set-up time and lack of haptic feedback have limited their use to expert centers only.

The concentric tubes approach is a third extracorporeal solution for robotizing the endoluminal surgery [13]. It aims at improving the motion control of catheters while reducing the overall size of the extracorporeal robotic system. This approach relies on a set of pre-shaped tubes translated and rotated into one-another by an external motorization unit. A proper combination of the tubes curvatures produces the desired motion at the catheter distal end. Application of concentric tubes to beating-heart intracardiac surgery has been investigated and in-vivo trials were performed [14]. However, the size and complexity of the extra-corporal drive unit is still important. Moreover, problems appear in controlling the distal tip due to complex transfer functions and stability issues [15]. Because the system is fully tele-operated, there is also a lake of force perception by the surgeon during the navigation process.

**Active catheters with embedded actuators**

For the aforementioned reasons, some authors have proposed the development of more ergonomic, low-cost and low-volume systems suited to easier daily usage. To this purpose, various solutions having the form of drivable active catheters with embedded micro-actuators were proposed in the literature.

Some of these solutions are based on electro-active polymers. Liu [16] imagined a micro-actuator based on an Ionic Conductive Polymer Film (ICPF) for controlling the bending of a tube. In [17], the electrochemical actuation of a catheter coated with polypyrrol is studied. This kind of actuator is interesting for its shapability and small volume but can't deliver the sufficient forces necessary to resist external contact in the blood vessels. Moreover there are issues about their biocompatibility. Another micro-technological solution relies on the hydraulic actuation principle. For example Ikuta [18] produced a hydrodynamic active catheter driven by integrated micro valves. Ikeuchi [19] proposes a pressure driven micro-catheter, made by laser ablation in a thermo-plastic membrane, with a 300 µm diameter. These solutions are interesting for safety and biocompatibility but rely on very complex manufacturing and assembling processes.

**SMA based active catheters**

Shape Memory Alloys (SMA) were also widely investigated in the field of active catheterism for their biocompatibility and excellent density-to-power ratio [20]. The basic concept of SMA actuation for catheterism is depicted on Fig.2. The bending motion is obtained by a mere electrical current $i$ which induces a phase transformation in the SMA material through the Joule effect. Here, the SMA actuator can have the shape of a flat spring or a coil spring or a wire. Besides, more than one SMA actuator are often integrated at the tip of the catheter in order to fully control its bending in the 3-D space (see Fig.3).

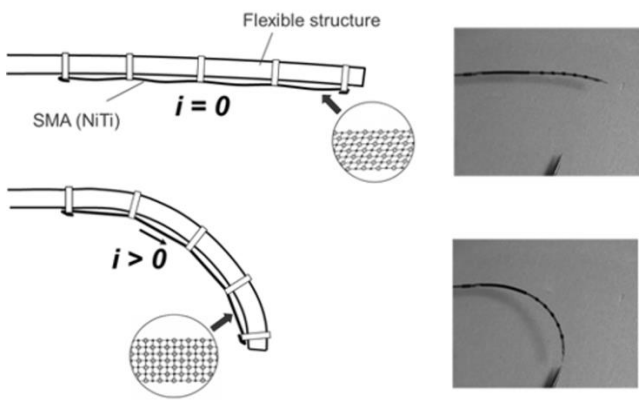

Fig. 2. SMA actuated catheter (principle) [21]

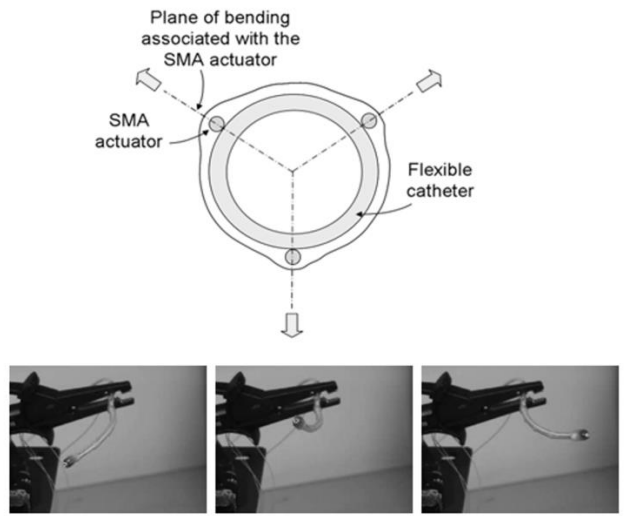

Fig. 3. Active catheter featured with three SMA actuators [21]

SMA actuators having the shape of flat springs have been used regarding the possibility to produce very small actuators [22–24]. In [23] for example, a 0.9-mm-diameter catheter has been realized this way. It integrates batch-fabricated small flat NiTi springs as actuators. The springs were obtained by photolithography and electrochemical etching. While kinematic performances with flat-spring SMA actuators are interesting (see Table 1), the fabrication process is always complex and thus poorly compatible with the low-cost criterion associated to single-use devices.

Table 1.  Performances of small dimeter active catheters using SMA actuators

| Authors | Ref | Year | Application field | Ext. diam. (mm) | Shape of the SMAs | Nb of SMA | Bending angle (deg) | Unit Length (mm) | Radius of curvature (mm) | Outer sheath | In-vivo tested |
|---------|-----|------|-------------------|-----------------|-------------------|-----------|---------------------|------------------|--------------------------|--------------|----------------|
| Haga | [28] | 1998 | Urology | 1.7 | Coil | 1 | 90 | 9 | 5.7 | no | no |
| Lim | [26] | 1995 | General | 2.8 | Coil | 3 | 8.5 | 5.2 | 35 | yes | no |
| Fu | [27] | 2008 | General | 1.3 | Coil | 3 | 90 | 30 | 19.2 | no | no |
| Mineta | [23] | 2001 | General | 0.9 | Flat | 3 | 35 | 12.4 | 20.3 | yes | no |
| Chang | [22] | 2002 | General | 3.0 | Flat | 3 | 80 | 40 | 28.7 | yes | no |
| Namazu | [24] | 2011 | General | 3.5 | Flat | 4 | 3 | 15 | 288 | yes | no |
| Szewczyk | [21] | 2011 | INR | 1.2 | Wire | 2 | 70 | 11 | 9.0 | no | no |
| Fukuda | [29] | 1994 | General | 2.0 | Wire | 3 | 80 | 40 | 28.6 | yes | yes |
| Mizuno | [31] | 1994 | Oncolog | 2.6 | Wire | 2 | 90 | 20 | 12.7 | yes | yes |
| Takizawa | [30] | 1999 | INR | 1.5 | Wire | 3 | 45 | 20 | 25.5 | yes | no |

Coil-shaped SMA micro-actuators were also used in active catheterism for their high strain capabilities [25-28]. However, coil-shaped SMA actuators are usually limited in the tensile force they can deliver. This leads to relatively poor kinematic performances as shown in Table 1. As we can see, in most cases coil-shaped SMA actuators result in high bending angles only at the expense of large radii of curvature. In [28], a small radius of curvature is achieved by an active catheter featured with only one coil-shaped SMA actuator. However, the proposed device is not covered by any external sheath or tube which is mandatory for safety purpose.

In contrary, wire-shaped SMA actuators exhibit high stress capabilities. They were thus often exploited in the design of active catheters. Their kinematic performances are shown in the last part of Table 1. Fukuda [29] and Takizawa [30] have developed catheters actuated by three SMA wires as depicted in Figure 3. These devices however bend with quite large radii of curvature. In [31], a similar device has been applied to peroral pancreatoscopy in pigs. It performs well in terms of curvature and bending angle but has a quite important outer diameter. In [21], the proposed catheter is quite thin and exhibits a small radius of curvature but has never been covered with any outer sheath and thus could not be tested in-vivo.

**Aim of the paper**

Our purpose is to develop a dexterous actuated catheter presenting low-cost, easy to use and small size properties. Consistently with the above state of the art, we directed our work toward the design of an active catheter embedding its own actuators. The embedded micro-actuators will be SMA wires for their simplicity of integration and kinematic performances. Moreover, only one SMA wire actuator will be embedded at a given bending section as the reachable radius of curvature seems to invariably decrease with the number of embedded actuators as illustrated on Fig.4.

In case of only one actuator at the catheter tip, we assume that the device can be manually rotated like a conventional catheter in order to reach any bending direction in the 3D space. Lastly, the core structure and assembly process of our device will not involve any complex operation in order to: a- limit the production costs, b- achieve very small outer diameters.

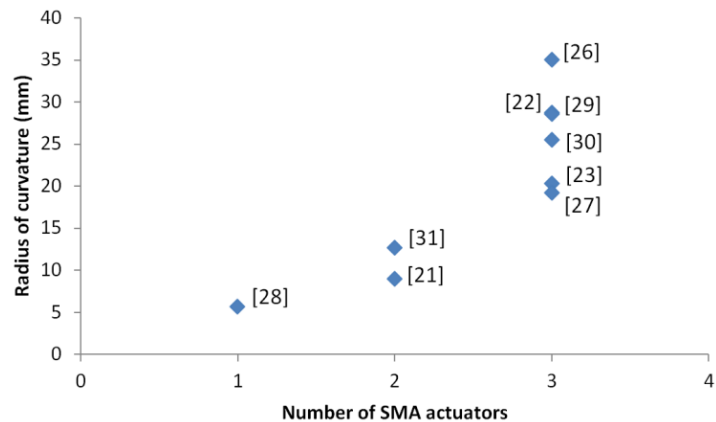

Fig. 4. Radius of curvature vs number of SMA actuators

In this paper, we first define functional and technical specifications for our device. Then we present the structure we developed and the model we used in order to optimize the sizing. In a second part, obtained kinematics performances are exposed and results of in-vitro and in-vivo experiments are finally given.

**MATERIAL AND METHODS**

**Anatomical considerations**

For practical reasons but without loss of generality, we focus in this paper on the cannulation of the peripheral arteries, namely: the supra-aortic trunks (brachiocephalic trunk, left common carotid artery and left subclavian artery), the celiac trunk, the superior mesenteric artery, the renal arteries and the common iliac arteries. Anatomical dimensions for these arteries were obtained from the review of the specialized literature [32-48]. They are listed in Tables 2 and 3.

Table 2.  Min and max diameters of the arteries of interest

| Zone | Vessel | Diameters [mm] | |
|---|---|---|---|
| | | Min | Max |
| Abdominal arota | Abdominal aorta | 15 | 20 |
| | Celiac trunk | 6 | 8 |
| | Common iliac artery | 7 | 10 |
| | Renal artery | 4 | 6 |
| Supra-aortic trunks | Distal aortic arch | 20 | 41 |
| | Brachiocephalic trunk | 11 | 31 |
| | Left common carotid | 7 | 14 |
| | Left subclavian artery | 5 | 18 |

Table 3.  Min and max bifurcation angles for the region of interest

| Zone | Vessel | Angulations [°] | |
|---|---|---|---|
| | | Min | Max |
| Abdominal arota | Abdominal aorta - celiac trunk | 115 | 155 |
| | Abdominal aorta - renal artery | 105 | 135 |
| | Abdominal aorta - superior mesenteric artery | 120 | 150 |
| Supra-aortic trunks | Aortic arch - brachiocephalic trunk | 27 | 96 |
| | Aortic arch - left common carotid | 5 | 97 |
| | Aortic arch - left subclavian artery | 2 | 102 |

**Functional specifications**

The functional specifications of our active bending device were derived from the above anatomical information. Our device is intended to be used in combination with a conventional 6F catheter (2mm internal diameter and 65cm length) which is the type of catheter generally used to navigate the above mentioned regions. This imposes to our device an outer diameter less than 2 mm and a total length greater than 65 cm.

We also relied on the existing material to specify the number of active bending sections. The most current approach for conventional catheters (e.g. "Renal double Curve" probe, "Cobra" probe) but also for steerable systems (e.g. Magellan®, Medrobotics Flex System®) is to combine two sequential curvatures in the same plane and with the same sense of bending. Our device will thus be featured with two sequentially arranged active bending sections. The first one is called the proximal section and the second one is called the distal section. These active sections will be separated by a passive section with minimal length $L_{sep}$ = 10 mm. This minimal spacing is necessary for assembly and connection purposes and is consistent with the values encountered on several existing devices [49].

In order to specify the lengths and the flexion angles of the two active sections we analyzed the way an operator should proceed to enter a lateral artery as illustrated on Fig. 5. The first step of the manœuver is to point toward the targeted artery by bending the distal section at 90° in the direction of the lateral ostium (see Fig. 5-b). For this reason, the total bending angle of the distal section must be at least $\theta_{2max}$ = 90°. To perform this movement without risking contacting and stucking against the arterial wall around the ostium, the corresponding radius of curvature $R_{2min}$ must be smaller than the diameter of the original artery. According to Table 3, the smaller aorta diameter is 15 mm which leads to $R_{2min}$ < 15mm.

The second step of the manœuver is to activate the proximal flexion to have the catheter extremity penetrate into the ostium with an alignment corresponding to the axis of the targeted artery (see Fig. 5-c). Taking into account that the maximal branch angulation reported in Table 4 is 155°, the maximal bending for the proximal flexion must be at least $\theta_{1max}$ = 65° in order to have : $\theta_{1max} + \theta_{2max} \geq 155°$.

Finally, in case of a very large original arteries, the operator has to cope with a large distance between the ostium and the opposite arterial wall (see Fig. 5-d). In this case, he/she will bend the proximal section while the distal one is kept straight. The aim here is to maximize the span $L_s$ of the device to be able to point and penetrate the targeted ostium despite its remoteness. Considering a maximal aorta diameter $D_{max}$ = 40 mm (see Table 3), we have to satisfy  $L_s > D_{max}$  (see Fig. 5-d), which is equivalent to:

$$( R_{1min} + L_{sep} + L_{2tot} ).\, sin(\,\theta_{1max}\,) >\, D_{max} \qquad (1)$$

where $R_{1min}$ is the radius of curvature of the proximal section when it is completely bent and $L_{2tot}$ stands for the total length of the distal active section.

The functional specifications of our device are summarized in Table 4.

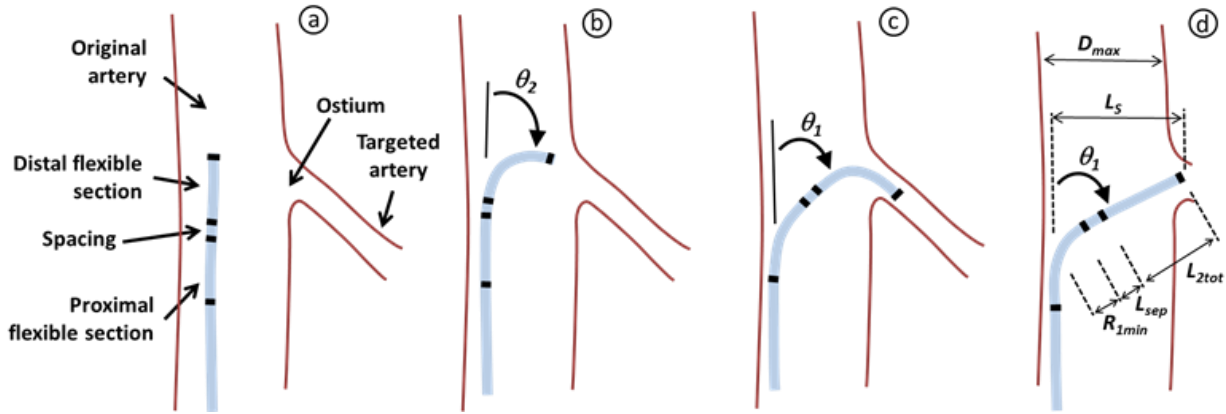

Fig. 5. Manœuver for entering a lateral artery: a- general description, b- pointing toward the ostium, c- entering into the ostium, d- case of a large original artery

Table 4. Functional specifications

| Constraints | Specification values |
| --- | --- |
| Total length of the device | > 650 mm |
| Outer diameter of the device | < 2 mm |
| Bending angle distal section $\theta_{2max}$ | > 90° |
| Bending angle proximal section $\theta_{1max}$ | > 65° |
| Radius of curvature distal section $R_{2min}$ | < 15 mm |
| Condition of reachability (inequality (1)) on $L_s$ | > 40 mm |

**Technical specifications**

The general description of our active device is given on Fig. 6 for a system comprised of two different active sections. All the mathematical symbols used in the following are listed in the nomenclature at the end of this paper.

As we can see on Fig. 6, any active section (either the proximal or the distal one) is composed of a flexible part and a rigid part. $L$ and $\theta_{max}$ are the length and the maximal bending angle of such a flexible part. And $R_{min}$ is the minimal radius of curvature associated to $L$ and $\theta_{max}$:

$$R_{min} = L / \theta_{max} \qquad (2)$$

In a given active section, the rigid part with length $L_{rig}$ is aimed at hosting the electrical connections as well as a portion of the associated SMA wire. Indeed, partly locating an SMA wire along a rigid section significantly improves its kinematic performances. In fact, assuming that an SMA wire undergoes a contraction rate $\tau_{max}$ in the fully bended configuration (Fig. 7), the radius of curvature of the bended part verifies:

$$(R_{min} - r)\theta_{max} = L - \tau_{max}.(L + L_{rig}) \quad (3)$$

where $r$ is the radial distance between the bended structure neutral line and the SMA axis (see Fig. 7). Finally, (2) combined with (1) leads to:

$$R_{min} = \frac{r \cdot L}{\tau_{max} \cdot (L + L_{rig})} \qquad (4)$$

From equation (4) it is obvious that including a rigid part with length $L_{rig}$ makes the radius of curvature smaller. We take advantage of this property throughout the development of our device.

**Core structure**

The core structure of an SMA-based active catheter is intended to host the SMA actuators as well as to produce an elastic restoring force to make the device naturally return to its straight configuration after activation. In our case, the core structure will be based on the combination of three thin metallic wires arranged in a triangle as shown on Fig.8. This approach has many advantages:

- to create the flexible part of a given active section, one of the three wires has just to be locally removed,

- the three metallic wires can be assembled in a stable triangular configuration just by ligating them without any gluing or soldering,

- the concavities appearing between the metallic wires help positioning the SMA wires exactly aligned with the device axis,

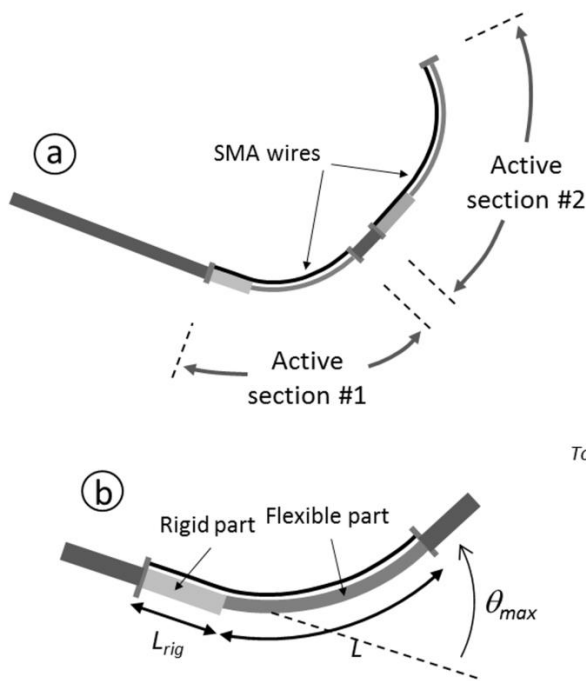

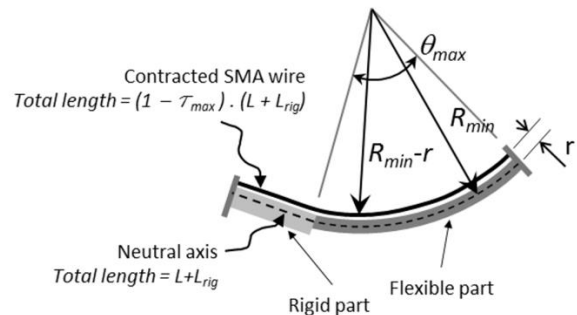

Fig. 6. Description of a device comprising two active sections; a- general description; b- detail of an active section

Fig. 7. Relation between the angle and the different radii of curvature

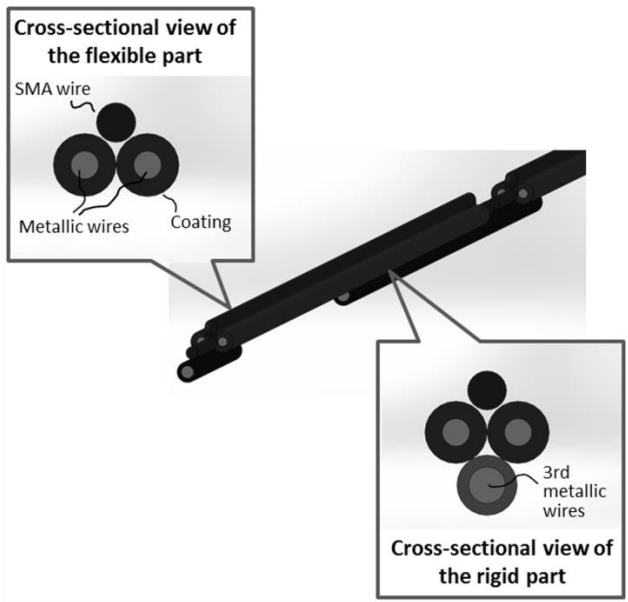

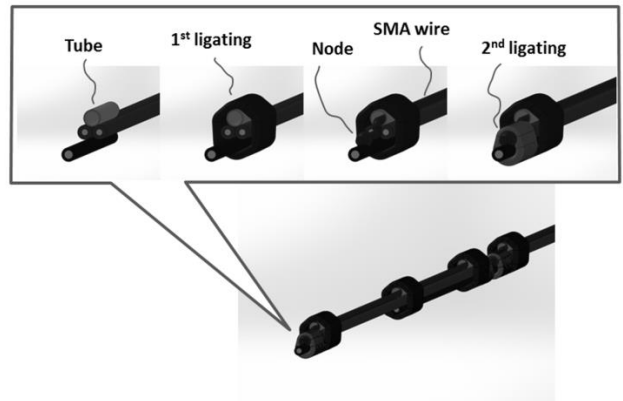

Fig. 8.  Cross-sectional views of
the flexible part and of the
rigid part of an active section

Fig. 9. Axial anchoring of the
SMA wire along the core
structure

Regarding the integration and the anchoring of the SMA wires on the structure, two

fundamental requirements have to be fulfilled:

1°) The longitudinal anchoring of an SMA wire has to be solid enough to resist the tensile forces generated during the flexion of the structure. In our case, we achieve this firm anchoring without any gluing or soldering but through the four different steps depicted on Fig.9.

2°) An SMA wire has to be maintained close to the bended flexible part of the structure in order to avoid any arch-like configuration. In our case, this is achieved by ligating a given SMA wire close to its corresponding flexible part using a large pitch helix made of a simple electrically isolated cupper wires (Fig. 10). This helix has a constant pitch of about 1 mm and doesn't need any peculiar fixation at its extremities.

Finally, to protect the active device we cover it by a common commercially available 6F catheter (Fig. 11).

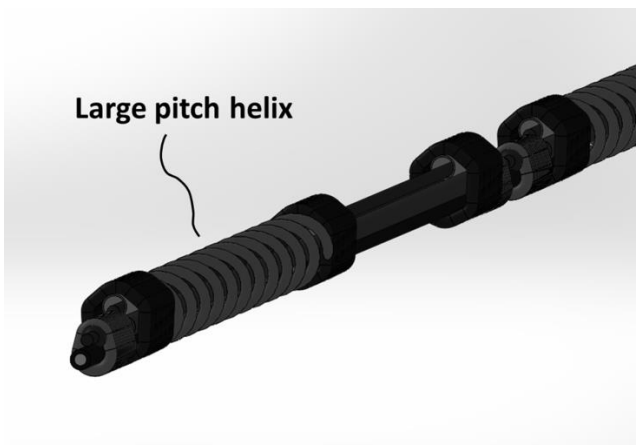

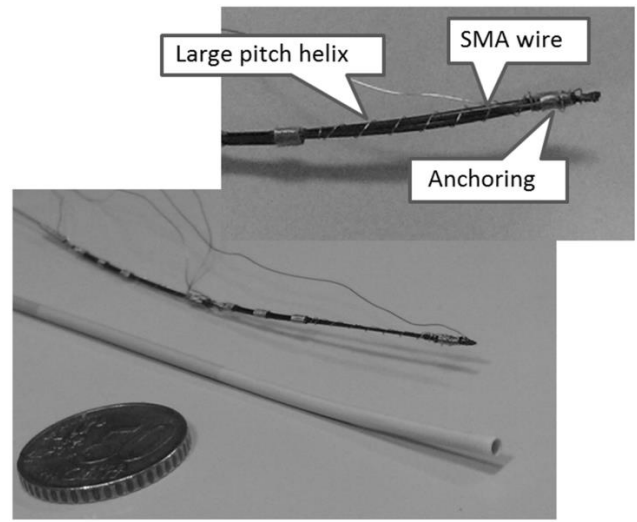

Fig. 10. Radial anchoring of the SMA wire

Fig. 11. The active device and its external covering catheter

**Sizing of the prototype**

In order to guide the sizing of our active catheter, we first derived a theoretical model of its kinemato-static behavior. According to this modelling (see the appendix section for the details of the derivation), the expression of the maximal bending angle of an active section is:

$$\theta_{max} = L \cdot \tau_{lim} \left( \frac{E_C(r_{ext}^4 - r_{int}^4) + 2.E_{met} \cdot r_{met}^4}{rE_A d^2} + \frac{rL}{L + L_{rig}} \right)^{-1} \quad (5)$$

The notations used in equ. (5) are illustrated on Fig. 12 and explained in the nomenclature that can be found at the end of this paper.

Regarding equ. (5), we can conclude that the maximum bending angle $\theta_{max}$ can be increased through the achievement of several physical compromises:

- The first term on the right side of equ. (5) indicates that the influence of the SMA wire diameter $d$ on $\theta_{max}$ is twofold : on the one hand it has to be as large as possible to increase this term and, on the other hand, it has to be kept small enough to cope with the maximal outside diameter specified in Table 4 (2 mm),

- Regarding the contradictory influence of the parameter $r$ on the two terms on the right hand side of equ. (5), there exists a theoretical optimal value for this parameter to be found,

- The diameter $r_{met}$ of the metallic wires should be chosen as small as possible regarding its influence on equ. (5) but should also be large enough to avoid any buckling phenomenon in the structure.

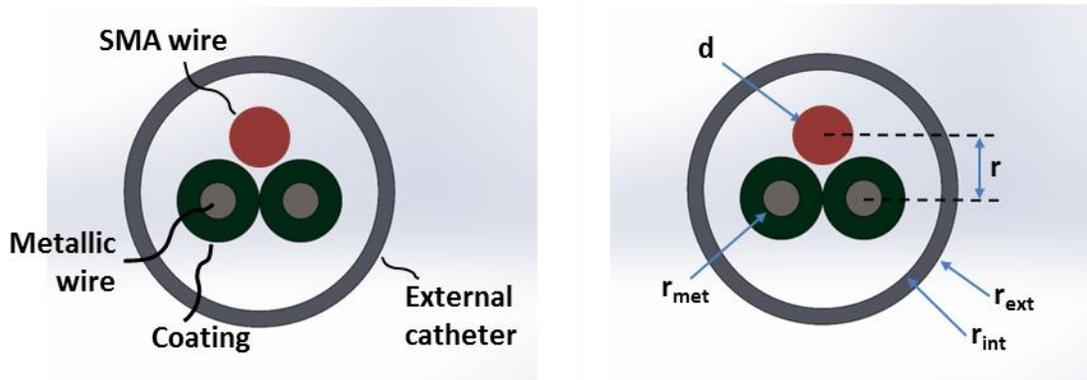

Fig. 12. Cross-sectional dimensions of the flexible part

Given the complexity of the above mathematical formulation, we chose not to rely on model (5) to determine the best dimensions for our device. We instead solved this optimization problem following a practical approach consisting in realizing and testing a large number of prototypes. Along this exploration, the geometrical parameters $d$, $r$, $r_{met}$, $L$ and $L_{rig}$ varied according to the values indicated in Table 5. The simplicity of the structure of our active catheter allowed us to build and test more than 70 prototypes. The mean time to build a complete prototype was about 2 hours. All the components of our active catheters were found among commercially available products. Lengths of the sections, radii of curvature and bending angles were graphically assessed for each prototype using images like Fig. 17. The estimated measuring errors are $\Delta\theta = 1°$, $\Delta L = 1$ mm and $\Delta R = 2$ mm. At the end of the process we selected the prototype with the best kinematic performances and whose characteristics best match the functional specifications of Table 4. The best prototype characteristics are given in Table 5 and are further discussed in the results section.

Besides to this dimensional optimization, we also investigated the validity of the theoretical model (5). For a randomized sample of 13 prototypes, we compared the experimental and theoretical values of the bending angles $\theta_{max}$ for both the proximal and the distal active sections.

Table 5.  Tested values of the geometric parameters

| Geometric Parameters | Symbols | Tested values | | Values of the best prototype |
|---|---|---|---|---|
| | | Min | Max | |
| SMA wire diameter | $d$ | 0.125 | 0.250 | 0.2 |
| Distance SMA axis – neutral line | $r$ | 0.1 | 0.2 | 0.2 |
| Metallic core radius | $r_{met}$ | 0.3 | 0.55 | 0.46 |
| Length of a flexible part | $L$ | 19 | 42 | 19 (prox.), 31.5 (dist.) |
| Length of a rigid part | $L_{rig}$ | 18 | 30 | 23 (prox.), 21 (dist.) |

**Experimental validation**

Our final prototype with its control interface were tested in-vitro and in-vivo. For the in-vitro experiments, we used a silicon-based, transparent, anthropomorphic phantom (Fig. 13) representing the thoracic and the abdominal aorta with their main branches (Elastrat Sarl, Geneva, Switzerland). A blood-mimicking mixture of water and glycerol (60%/40%) was circulated in the model with a pump. We tried out the cannulation of the contralateral common iliac artery, of both renal arteries (with different angles: 90°, 120° and 150°), and of all the supra-aortic trunks by a right femoral approach. Success or failure to cannulate was recorded for each artery.

In-vivo experiments were carried out on a 23 kg pig under general anaesthesia and fluoroscopic guidance in the operating room of the Ecole de Chirurgie (UPMC, Paris, France). Through a right femoral approach, we tested the cannulation of the supra-aortic trunks, of both renal arteries and of the celiac trunk. We systematically performed an angiographic control after each cannulation in order to look for an embolism or a perforation.

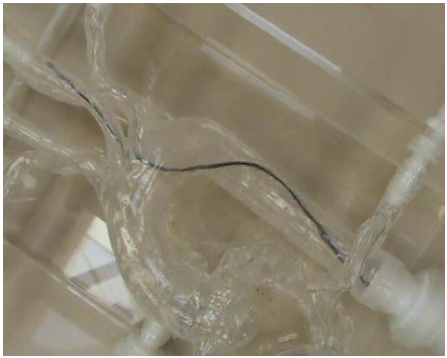

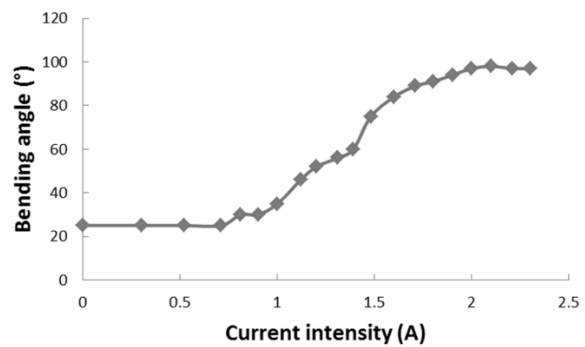

Fig. 13. Test on a silicon-based anatomical phantom

Fig. 14. Transfer function between the current intensity $i$ and the bending angle $\theta$

**Control architecture**

The SMA actuators are activated in an open-loop manner relying on the transfer function between the supplied currents $i$ in a SMA actuator and the resulting bending of the activated section of the catheter (Fig. 14). At a higher level of control, the operator is adjusting the catheter configuration through a visual feedback featured by a X-ray imaging system. As we can see on Fig. 14, a minimal current intensity (and voltage) of approximately 0,8 A (and 4 V) is required to start the bending of the flexible part. Besides, there is a maximal current intensity and voltage (2 A and 10 V) beyond which no additional bending is observed. Between these two values, the angulation of the flexible part of the catheter is nearly linear with respect to $i$. This allowed us to implement a linear feedforward controller.

The hardware architecture of our control solution is depicted on Fig.15. An Arduino Uno card with an ATmega328P microcontroller (Atmel Corporation, San Jose, CA) was used to host the program converting the command from the operator into an electric signal. The power amplification is achieved using a 4x4 Driver Shield (Logos Electromechanical, Seattle, WA). We found that a minimal PWM period of 20 ms was necessary to avoid vibrations of the active device. We chose a pedal board as human-machine interface (Fig. 16). This kind of interface let the operator handle the catheter and the guidewire with his/her hands as usual.

**RESULTS**

**Dimensioning of the active catheter**

Our final prototype is depicted on Fig.17. The external catheter is an Envoy 6F catheter (Codman Neuro, Raynham, MA). The metallic wires are Radifocus Guidewires M Standard Type (Terumo France SAS, Guyancourt, France) having a diameter of 0,018 inches (= 0.46 mm). The SMA wires are SmartFlex wires (Memry GmbH, Weil am Rhein,

Germany) with diameter 0.2 mm. Their lengths are: $L$ = 24.5 mm and $L_{rig}$ = 21 mm for the first active section and $L$ = 19 mm and $L_{rig}$ = 23 mm for the second active section. The tip of the device is composed of a 5 mm long rigid portion used for various SMA anchoring and electrical connections.

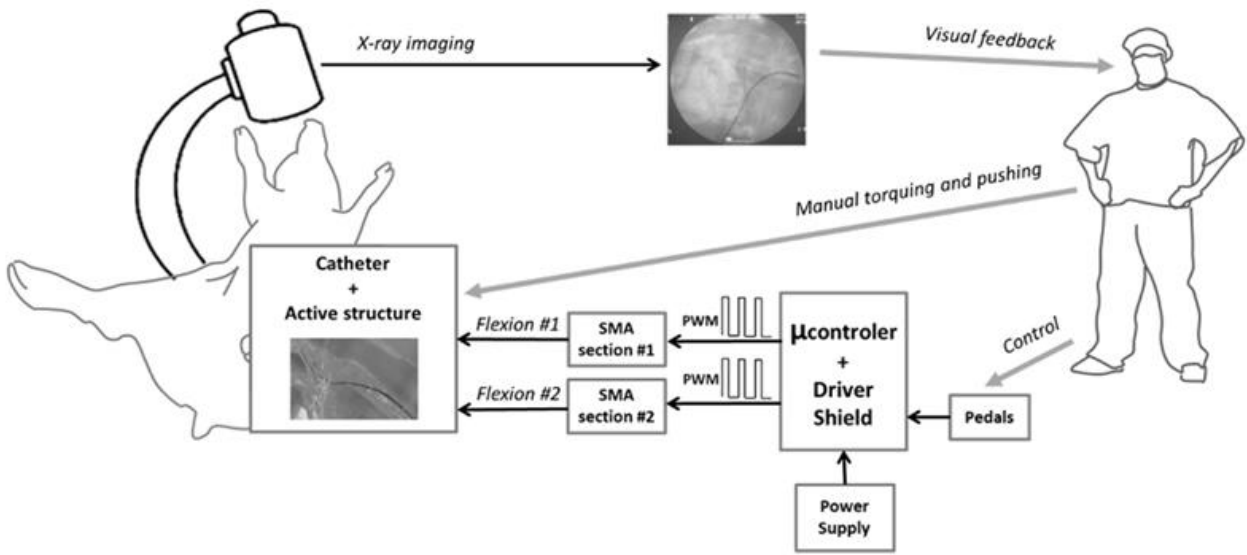

Fig.15 Experimental set-up

Table 6 presents the functional characteristics of our best prototype compared to the initial specifications. Length, diameters and bending angles satisfy the initial specifications. Only the bending radius of the distal section doesn't exactly comply with the corresponding required value. This may lead to unexpected contacts between the device tip and the arterial wall especially when pointing to a lateral ostium located in a small diameter section of the abdominal aorta (e.g. : the renal arteries or the celiac trunk as can be seen in Table 2). However, given the relatively small difference between the desired and actual radii of curvature, the selected prototype appears to be acceptable regarding its global kinematic performances.

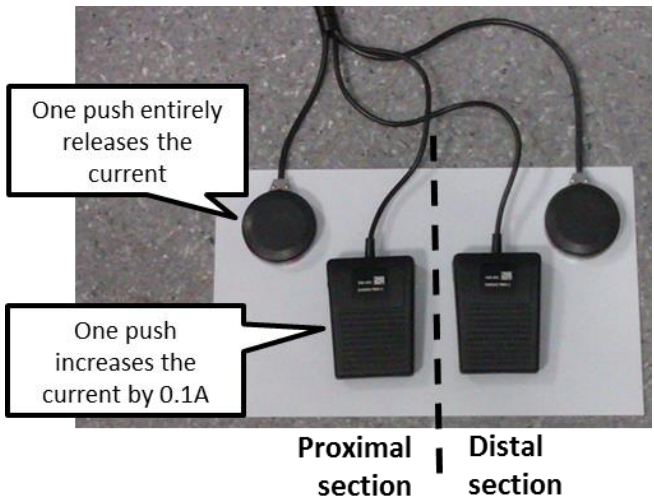

Fig. 16. Control pedal used for in-vitro and in-vivo experiments

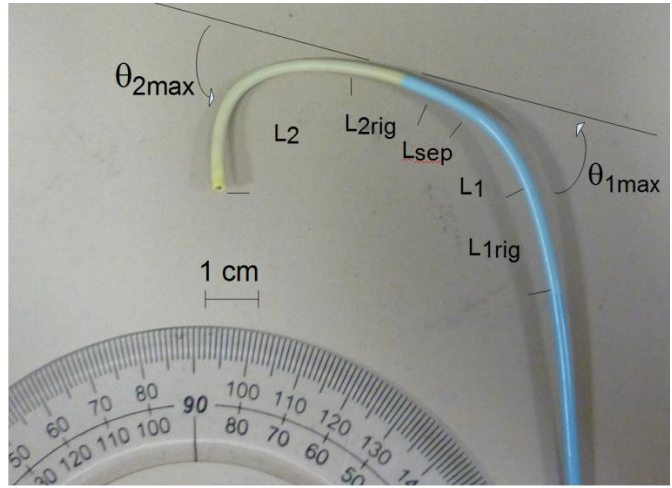

Fig. 17. Final prototype in maximal bending conditions

On Fig. 18, we represented the evolution of the maximal bending angle of the distal section of a prototype having the dimensions specified in Table 5. We can see that the performances are stable until 280 cycles which is enough to perform an even complex surgical procedure. During this experiment, we used an activation current of 2A and the device, surrounded by an Envoy 6F catheter, was immersed in a 37°C water-bath. The catheter external temperature showed no significant change during this test.

Finally, of the 70 tested prototypes 13 were randomly selected in order to evaluate the relevance of the theoretical model (equ.(5)) in terms of $\theta_{max}$ prediction. In equ. (5), the catheter Young modulus was fixed to $E_c$ = 1.9 GPa according to the values that can be found in the literature [38] and to the composition of the Envoy 6F catheter we used : nylon ( Young's modulus $E$ = 2 GPa), polyurethane ($E$ = 1,5 GPa), stainless steel ($E$ = 220 GPa) in minority, PTFE ($E$ = 0,5 GPa). Fig. 19 represents the relative errors between the experimental and theoretical values of $\theta_{max}$.

Table 6.  Comparison between the initial functional specifications
and the characteristics of the best prototype

| Constraints | Specification values | Best prototype values |
| --- | --- | --- |
| Total length of the device | > 650 mm | **1000 mm** |
| Outer diameter of the device | < 2 mm | **0.98 mm** |
| Bending angle distal section $\theta_{2max}$ | > 90° | **110°** |
| Bending angle proximal section $\theta_{1max}$ | > 65° | **70°** |
| Radius of curvature distal section $R_{2min}$ | < 15 mm | **16.4 mm** |
| Condition of reachability (inequality (1)) $L_s$ | > 40 mm | **74.1 mm** |

**Experimental validation**

The experimental validation was performed on the silicon model described earlier. The active catheter was able to cannulate all the supra-aortic trunks (brachiocephalic trunk, left common carotid artery and left subclavian artery) and both renal arteries (with different angles: 90°, 120° and 150°) via a right femoral approach. The catheterization of the contralateral common iliac artery failed and didn't allow the catheter tip to enter into the ostium of the considered artery. The total span of the prototype when bended at its maximal capabilities was too important. The main reason for that is a slight unexpected bending of the supposed rigid part of the distal active section.

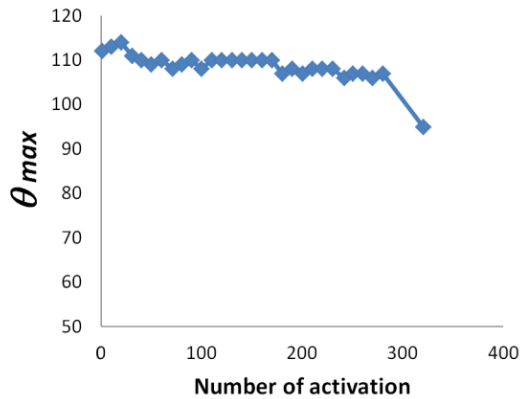

Fig. 18. Evolution of $\theta_{max}$ with respect to the number of activations of the SMA (distal segment)

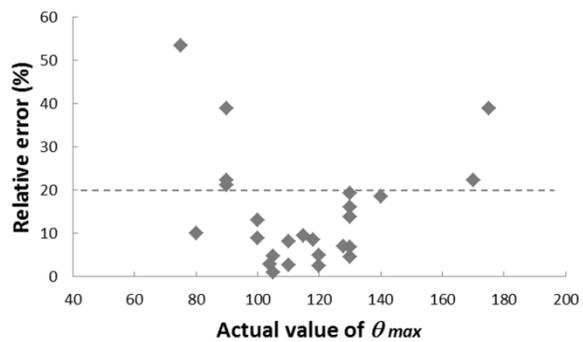

Fig. 19. Accuracy of the theoretical model in predicting $\theta_{max}$

Throughout our in-vivo experiments on a 23-kg pig (Fig. 20), all the cannulations were successful: both renal arteries, the celiac trunk, all the supra-aortic trunks (common trunk for the right subclavian artery and both common carotid arteries). The angle between the aorta and the celiac trunk was 45° and the angles between renal arteries and the aorta were 80° and 90°. The pig however was too small to cannulate the contralateral iliac. On the other hand, because of the small size of the arteries, this experiment allowed us to evaluate the feasibility and safety of navigation in a difficult anatomy. Security was evaluated after each cannulation. All the arteriographies showed integrity of the arterial wall (neither perforation nor dissection) – no emboli or thrombosis were visualized. Some post-cannulation arteriographies and successful catheterizations are presented in Fig. 21. Five copies of the active catheter sized according to the results reported in Table 5 were used during this experiment. We didn't find any evidence of thrombosis inside any of these catheters. Each prototype has been used for a mean duration of 30 minutes, with a mean time of activation of 20 minutes.

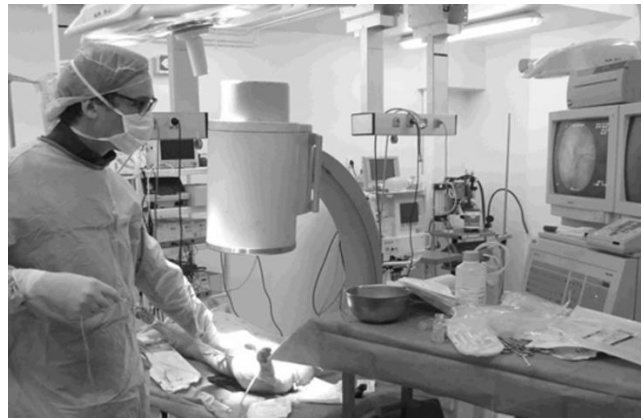

Fig. 20. Experimental set up for in vivo validation

**DISCUSSION**

Active catheters based on the SMA wire actuation principle have many advantages over cable-driven catheters in terms of miniaturization, accuracy of movement and number of controllable degrees of freedom. We determined the functional specifications of our device based upon a quantitative analysis of anatomical data and a simple qualitative kinematic reasoning. The specifications we found are close to the dimensions of the Magellan robot which is the only clinically-in-use robot for peripheral arteries. The active catheter we developed has a very simple and low-cost mechatronical structure. Only 2 hours are needed to build a complete prototype.

In terms of dimensions and kinematic performances, our final device compares favorably to the active catheters of the literature presented in Table 1. Compared to the device presented in [31], which appears to be the best existing prototype among those in Table 1 that are equipped with an external protective sheath, our final device shows a quite similar radius of curvature for an equivalent external diameter and with a significantly higher maximal bending angle.

We performed in-vitro and in-vivo experiments to validate the robustness and safety of our active catheters. We were able to cannulate several bifurcations up to 150° in a silicon-based model but failed to pass the aortic bifurcation because our device had a too large span regarding the diameter of the distal aorta. Throughout an experiment on animal, we were able to navigate through the peripheral arteries without causing any damage to the vessels (Fig. 21). Note that a glucose flush between the active structure and the external catheter was maintained during the experiment in order to limit temperature elevation. On a pig having a narrow aorta, we successfully cannulated the supra-aortic trunks, the celiac trunk and both renal arteries. We faced some difficulties in performing the cross-over (go into the contralateral iliac artery). These difficulties are probably due to an unexpected bending of the supposed rigid parts during activation of the catheter.

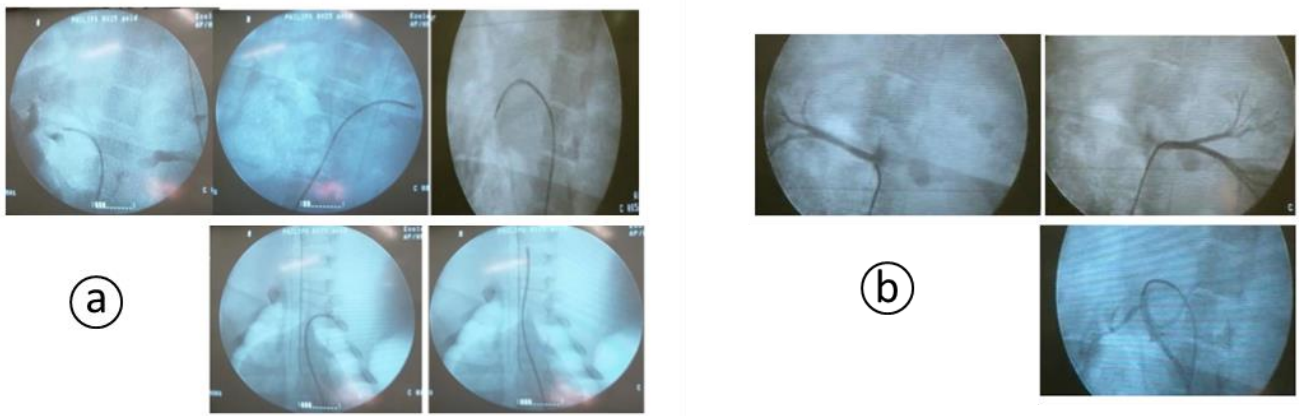

Fig. 21. a- Successful cannulations of right renal artery, left renal artery, celiac trunk, left subclavian artery, common bi-carotid and right subclavian trunk; b- angiographic control after cannulation of the right and left renal arteries, and of the celiac trunk

In comparison, Clogenson [50] could reach the contralateral iliac bifurcation, the left renal artery, the left subclavian artery and the brachiocephalic trunk on the Elastrat model with their multiselective MRI-guided catheter. But on the animal they only could cannulate the supraaortic trunks because the abdominal aorta was too narrow for their prototype. In another publication, the authors could successfully cannulate most of the small bifurcations with an angle inferior to 120°, but their deflectable guidewire was unable to cross bifurcations with large diameters [32]. Besides, only two studies to our knowledge have described the use of SMA-based catheters during in-vivo experiments. Fukuda have evaluated an active catheter on a dog but no explicit results were given [29]. Mizuno tested its active catheters on 19 patients with good results but the field of application was the peroral pancreatoscopy and thus was completely different from ours [31]. In particular, thermal effects in hollow organs have less probability to appear than in blood.

In this paper, we also developed a mathematical model for predicting the bending angle and radius of curvature of our active catheters. We first used this model to state some general qualitative rules useful for the iterative dimensional optimization of our devices. We then made a quantitative comparison between the actual and predicted bending angles for a sample of prototypes. It happens that the relative error is less than 20% for bending angles between 100° and 150° which is the interval of interest for our applications. Based on this model, a more formal and systematic dimensioning process can now be developed for serving future applications. For example, the theoretical variations of $\theta_{max}$ with respect to $r$ (Fig.22) can be used to adjust the diameter of the loose ligatures and maximize the range of bending of our devices.

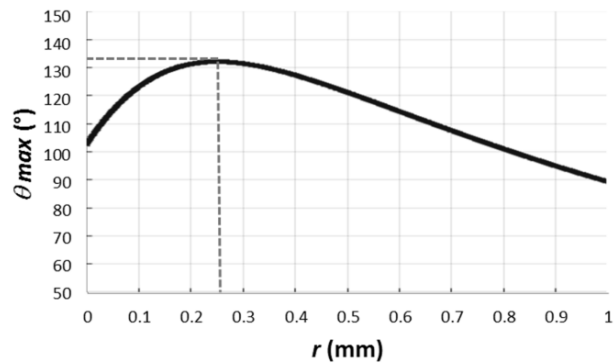

Fig. 22. Expression of θmax in function of r
according to the theoretical model

**CONCLUSION**

Endovascular treatment of arterial pathologies using conventional material is time consuming and faces major difficulties in its access phase due to anatomical constraints. In this paper we propose a new kind of SMA-based active catheter to facilitate the endovascular navigation. The new SMA-based catheter we present has a core structure made of three thin metallic wires in triangle. Such a structure allows an easy anchoring of the SMA wires and a precise control of the bending curve. A 6F introductor can be used as an external protective sheath thanks to the limited overall diameter of our core structure which is less than 2mm.

The technical specifications of our device were identified relying on the specialized literature. A trial-error campaign comparing about 70 different prototypes was conducted to determine the best dimensions of the core structure and of the SMA actuators with respect to the imposed specifications. A second major contribution of this paper is the derivation of a reliable mathematical model for predicting the kinematic behavior of such an active catheter.

In-vitro experiments in realistic conditions proved the validity of the concept for most of the peripheral arteries and for different bifurcation angles. Several prototypes were also extensively tested in-vivo on a 23 kg pig. We succeeded in the cannulation of numerous peripheral arteries without damaging any of the arterial walls or provoking any emboli or thrombosis. To our knowledge, this is the first time an SMA-based active catheter successfully performs the cannulation of several peripheral arteries in a real operating environment.

Our next study aims at evaluating the relevancy of our active catheters in a more statistical manner. We intend to conduct a quantitative campaign to assess the benefit of our active catheters in difficult contexts such as the catheterism of stenosed arteries and the crossing of tight bifurcation angles. This campaign will be conducted on anatomical phantoms and will involve experienced surgeons as well as young fellows. In a future work we also want to extend our approach to other endovascular technics such as interventional neuroradiology. The very simple structural composition of our device will be preserved and exploited to conduct a size scale reduction accounting for the tiny diameters of the brain vessels. To operate this scale reduction we will also rely on the theoretical model (5) validated in this paper.

**COMPLIANCE WITH ETHICAL STANDARDS**

Thibault Couture and Jérôme Szewczyk certify that they have no affiliations with or involvement in any organization or entity with any financial interest in the subject matter or materials discussed in this manuscript. All procedures performed on animal were in accordance with the ethical standards of the ethics committee of Paris 6 University Hospital. Appropriate care was taken to avoid any psychological or physiological distress of the animal.

**FUNDING**

This work was supported by French state funds managed by the ANR within the Investissements d'Avenir program (Labex CAMI) under reference ANR-11-LABX-0004 and by the Association chirurgicale pour le développement et l'amélioration des techniques de dépistage et de traitement des maladies cardio-vasculaires (ADETEC).

**NOMENCLATURE**

| | |
|---|---|
| $\theta_{max}$ | maximum bending angle of an active section |
| $\tau_{max}$ | contraction rate of an SMA wire when $\vartheta_{max}$ is reached |
| $\tau_{lim}$ | contraction rate of an SMA wire in a fully austenitic state |

and without any external load

| | |
|---|---|
| $d$ | diameter of the SMA wire |
| $r$ | radial distance between the neutral line and the SMA axis |
| $r_{ext}$ | external radius of the catheter |
| $r_{int}$ | internal radius of the catheter |
| $r_{met}$ | radius of the metallic core in the guide wires |
| $D_{max}$ | maximum aorta diameter |
| $E_A$ | Young modulus of the SMA material |
| $E_c$ | Young modulus of the catheter |
| $E_{met}$ | Young modulus of the metallic core in the guide wires |
| $I_c$ | quadratic moment of the catheter |
| $I_{met}$ | quadratic moment of the metallic core in the guide wires |
| $L$ | lenght of the flexible part of an active section |
| $L_{rig}$ | length of the rigid part of an active section |
| $L_{tot}$ | total length of an active section |
| $L_{sep}$ | spacing between two active sections |
| $M$ | distributed moment along the bended structure |
| $R_{min}$ | minimum radius of curvature of an active section |
| $S_A$ | minimum radius of curvature of an active |
| $T$ | tensile strength undergone by the activated SMA wire |

**REFERENCES**

[1] Coghlan KM, Breen LT, Martin Z, O'Neill S, Madhaven P, Moore D, Murphy BP (2013) An experimental study to determine the optimal access route for renal artery interventions. *Eur J Vasc Endovasc Surg Off J Eur Soc Vasc Surg* 46:236–241. doi: 10.1016/j.ejvs.2013.05.001

[2] Antoniou GA, Riga CV, Mayer EK, Cheshire NJ, Bicknell CD (2011) Clinical applications of robotic technology in vascular and endovascular surgery. *J Vasc Surg* 53:493–499. doi: 10.1016/j.jvs.2010.06.154

[3] Cochennec F, Riga C, Hamady M, Cheshire N, Bicknell C (2013) Improved catheter navigation with 3D electromagnetic guidance. *J Endovasc Ther Off J Int Soc Endovasc Spec* 20:39–47. doi: 10.1583/12-3951.1

[4] Duran C, Lumsden AB, Bismuth J (2014) A randomized, controlled animal trial demonstrating the feasibility and safety of the Magellan$^{TM}$ endovascular robotic system. *Ann Vasc Surg* 28:470–478. doi: 10.1016/j.avsg.2013.07.010

[5] Bismuth J, Kashef E, Cheshire N, Lumsden AB (2011) Feasibility and safety of remote endovascular catheter navigation in a porcine model. *J Endovasc Ther Off J Int Soc Endovasc Spec* 18:243–249. doi: 10.1583/10-3324R.1

[6] Bismuth J, Duran C, Stankovic M, Gersak B, Lumsden AB (2013) A first-in-man study of the role of flexible robotics in overcoming navigation challenges in the iliofemoral arteries. *J Vasc Surg* 57:14S–9S. doi: 10.1016/j.jvs.2012.08.124

[7] Cochennec F, Kobeiter H, Gohel M, Marzelle J, Desgranges P, Allaire E, Becquemin JP (2015) Feasibility and safety of renal and visceral target vessel cannulation using robotically steerable catheters during complex endovascular aortic procedures. *J Endovasc Ther Off J Int Soc Endovasc Spec* 22:187–193. doi: 10.1177/1526602815573228

[8] Riga CV, Cheshire NJW, Hamady MS, Bicknell CD (2010) The role of robotic endovascular catheters in fenestrated stent grafting. *J Vasc Surg* 51:810–819; discussion 819–820. doi: 10.1016/j.jvs.2009.08.101

[9] Riga CV, Bicknell CD, Hamady MS, Cheshire NJW (2011) Evaluation of robotic endovascular catheters for arch vessel cannulation. *J Vasc Surg* 54:799–809. doi: 10.1016/j.jvs.2011.03.218

[10] Riga CV, Bicknell CD, Hamady M, Cheshire N (2012) Tortuous iliac systems - a significant burden to conventional cannulation in the visceral segment: is there a role for robotic catheter technology? *J Vasc Interv Radiol JVIR* 23:1369–1375. doi: 10.1016/j.jvir.2012.07.006

[11] de Ruiter QMB, Moll FL, van Herwaarden JA (2015) Current state in tracking and robotic navigation systems for application in endovascular aortic aneurysm repair. *J Vasc Surg* 61:256–264. doi: 10.1016/j.jvs.2014.08.069

[12] Shurrab M, Danon A, Lashevsky I, Kiss A, Newman D, Szili-Torok T, Crystal E (2013) Robotically assisted ablation of
atrial fibrillation: a systematic review and meta-analysis. *Int J Cardiol* 169:157–165. doi: 10.1016/j.ijcard.2013.08.086

[13] Gilbert HB, Hendrick RJ, Webster RJ (2016) Elastic Stability of Concentric Tube Robots: A Stability Measure and Design Test. *IEEE Trans Robot Publ IEEE Robot Autom Soc* 32:20–35. doi: 10.1109/TRO.2015.2500422

[14] Dupont PE, Gosline A, Vasilyev N, Del Nido P (2012) Concentric Tube Robots for Minimally Invasive Surgery. *ResearchGate*

[15] Kim JS, Lee DY, Kim K (2014) Toward a solution to the snapping problem in a concentric-tube continuum robot: Grooved tubes with anisotropy. In: *2014 IEEE Int. Conf. Robot. Autom. ICRA*. pp 5871–5876

[16] Liu J, Wang Y, Zhao D, Zhang C, Chen H, Li D (2014) Design and fabrication of an IPMC-embedded tube for minimally invasive surgery applications. *Proceeding of SPIE* p 90563K–90563K–9

[17] Shoa T, Madden JD, Fekri N, Munce NR, Yang VX (2008) Conducting polymer based active catheter for minimally invasive interventions inside arteries. *Conf Proc Annu Int Conf IEEE Eng Med Biol Soc IEEE Eng Med Biol Soc Annu Conf* 2008:2063–2066. doi: 10.1109/IEMBS.2008.4649598

[18] Ikuta K, Yajima D, Ichikawa H, Katsuya S (2007) Hydrodynamic Active Catheter with Multi Degrees of Freedom Motion. In: Magjarevic R, Nagel JH (eds) *World Congr. Med. Phys. Biomed. Eng.* 2006. Springer Berlin Heidelberg, pp 3091–3094

[19] Ikeuchi M, Ikuta K (2008) Membrane micro emboss following excimer laser ablation

(MeME-X) process for pressure-driven micro active catheter. *2009 IEEE Int. Conf. Robot. Autom. ICRA*, pp 62–65

[20] Huber JE, Fleck NA, Ashby MF (1997) The selection of mechanical actuators based on performance indices. *Proc R Soc Lond Math Phys Eng Sci* 453:2185–2205. doi: 10.1098/rspa.1997.0117

[21] Szewczyk J, Marchandise E, Flaud P, Royon L, Blanc R (2011) Active Catheters for Neuroradiology. *J Robot Mechatron* 23:105–115. doi: 10.20965/jrm.2011.p0105

[22] Chang JK, Chung S, Lee Y, Park J, Lee SK, Yang SK, Moon SY, Tschepe J, Chee Y, Han DC (2002) Development of endovascular microtools. *J Micromechanics Microengineering* 12:824. doi: 10.1088/0960-1317/12/6/313

[23] Mineta T, Mitsui T, Watanabe Y, Kobayashi S, Haga Y, Esashi M. (2001) Batch fabricated flat meandering shape memory alloy actuator for active catheter. *Sens Actuators Phys* 88:112–120. doi: 10.1016/S0924-4247(00)00510-0

[24] Namazu T, Komatsubara M, Nagasawa H, Miki T, Tsurui T, Inoue S (2011), Titanium-Nickel Shape Memory Alloy Spring Actuator for Forward-Looking Active Catheter. *J Metall 2011*, 2011:e685429. doi: 10.1155/2011/685429, 10.1155/2011/685429

[25] Haga Y, Esashi M, Maeda S (2000) Bending, torsional and extending active catheter assembled using electroplating. In: *Thirteen. Annu. Int. Conf. Micro Electro Mech. Syst*. 2000 MEMS 2000. pp 181–186

[26] Lim G, Minami K, Sugihara M (1995) Active catheter with multi-link structure based on silicon micromachining. In: *IEEE Micro Electro Mech. Syst. 1995 MEMS 95 Proc.* p 116

[27] Fu Y, Li X, Wang S, Liu H, Liang Z (2008) Research on the axis shape of an active catheter. *Int J Med Robot* 4:69–76. doi: 10.1002/rcs.172

[28] Haga Y, Tanahashi Y, Esashi M (1998) Small diameter active catheter using shape memory alloy. In: *Elev. Annu. Int. Workshop Micro Electro Mech. Syst. 1998 MEMS 98 Proc.* pp 419–424

[29] Fukuda T, Guo S, Kosuge K (1994) Micro active catheter system with multi degrees of freedom. In: *1994 IEEE Int. Conf. Robot. Autom. 1994 Proc.* pp 2290–2295 vol.3

[30] Takizawa H, Tosaka H, Ohta R, Kaneko S, Ueda Y (1999) Development of a microfine

active bending catheter equipped with MIF tactile sensors. In: *Twelfth IEEE Int. Conf. Micro Electro Mech. Syst. 1999 MEMS 99*. pp 412–417

[31] Mizuno S, Nakajima M, Yasuda K, Kobayashi M, Mukai H, Hirano S, Kawai K (1994) Shape memory alloy catheter system for peroral pancreatoscopy using an ultrathin-caliber endoscope. *Endoscopy* 26:676–680. doi: 10.1055/s-2007-1009064

[32] Clogenson HCM, Simonetto A, van den Dobbelsteen JJ (2015) Design optimization of a deflectable guidewire. *Med Eng Phys* 37:138–144. doi: 10.1016/j.medengphy.2014.10.004

[33] Wilbring M, Rehm M, Ghazy T, Amler M, Matschke K, Kappert U (2016) Aortic Arch Mapping by Computed Tomography for Actual Anatomic Studies in Times of Emerging Endovascular Therapies. *Ann Vasc Surg* 30:181–191. doi: 10.1016/j.avsg.2015.07.018

[34] Vučurević G, Marinković S, Puškaš L, Kovačević I, Tanasković S, Radak D, Ilić A (2013) Anatomy and radiology of the variations of aortic arch branches in 1,266 patients. *Folia Morphol* 72:113–122.

[35] Demertzis S, Hurni S, Stalder M, Gahl B, Herrmann G, Van den Berg J (2010) Aortic arch morphometry in living humans. *J Anat* 217:588–596. doi: 10.1111/j.1469-7580.2010.01297.x

[36] Chiu P, Lee HP, Venkatesh SK, Ho P (2013) Anatomical characteristics of the thoracic aortic arch in an Asian population. *Asian Cardiovasc Thorac Ann* 21:151–159. doi: 10.1177/0218492312449637

[37] Finlay A, Johnson M, Forbes TL (2012) Surgically relevant aortic arch mapping using computed tomography. *Ann Vasc Surg* 26:483–490. doi: 10.1016/j.avsg.2011.08.018

[38] Shin I-Y, Chung Y-G, Shin W-H, Im S-B, Hwang S-C, Kim B-T, (2008) A morphometric study on cadaveric aortic arch and its major branches in 25 korean adults : the perspective of endovascular surgery. *J Korean Neurosurg Soc* 44:78–83. doi: 10.3340/jkns.2008.44.2.78

[39] Terumo AZUR peripheral hydrocoil embolisation system. 2014. Average vessel sizes and choice of the suitable coil size. *Available at: http://www.terumois. com/administration/Collateral.aspx*.

[40] Noordergraaf A, Verdouw D, Boom HB (1963) The use of an analog computer in a

circulation model. *Prog Cardiovasc Dis* 5:419–439.

[41] Kahraman H, Ozaydin M, Varol E, Aslan SM, Dogan A, Altinbas A, Demir M, Gedikli O, Acar G, Ergene O (2006) The diameters of the aorta and its major branches in patients with isolated coronary artery ectasia. *Tex Heart Inst J* 33:463–468.

[42] Turba UC, Uflacker R, Bozlar U, Hagspiel KD (2009) Normal renal arterial anatomy assessed by multidetector CT angiography: are there differences between men and women? *Clin Anat* N Y N 22:236–242. doi: 10.1002/ca.20748

[43] Kaufman J, Lee M (2004) Vascular and Interventional Radiology: the requisites. *Elsevier*

[44] Rogers IS, Massaro JM, Truong QA, Mahabadi AA, Kriegel MF, Fox CS, Thanassoulis G, Isselbacher EM, Hoffmann U, O'Donnell CJ (2013) Distribution, determinants, and normal reference values of thoracic and abdominal aortic diameters by computed tomography (from the Framingham Heart Study). *Am J Cardiol* 111:1510–1516. doi: 10.1016/j.amjcard.2013.01.306

[45] Pedersen OM, Aslaksen A, Vik-Mo H (1993) Ultrasound measurement of the luminal diameter of the abdominal aorta and iliac arteries in patients without vascular disease. *J Vasc Surg* 17:596–601.

[46] Shah PM, Scarton HA, Tsapogas MJ (1978) Geometric anatomy of the aortic--common iliac bifurcation. *J Anat* 126:451–458.

[47] Silveira LA da, Silveira FBC, Fazan VPS (2009) Arterial diameter of the celiac trunk and its branches. Anatomical study. *Acta Cirúrgica Bras Soc Bras Para Desenvolv Pesqui Em Cir* 24:43–47.

[48] Malnar D, Klasan GS, Miletić D, Bajek S, Vranić TS, Arbanas J, Bobinac D, Coklo M (2010) Properties of the celiac trunk--anatomical study. *Coll Antropol* 34:917–921.

[49] Helene C.M. Clogenson, Joris Y. van Lith, Jenny Dankelman, Andreas Melzer, John J. van den Dobbelsteen. Multi-selective catheter for MR-guided endovascular interventions. Medical Engineering and Physics 000 (2015) 1–8

[50] Clogenson HCM, van Lith JY, Dankelman J, Melzer A, van den Dobbelsteen JJ (2015) Multi-selective catheter for MR-guided endovascular interventions. *Med Eng Phys* 37:623–630. doi: 10.1016/j.medengphy.2015.03.015

**APPENDIX : THEORETICAL MODEL OF A SMA-BASED FLEXIBLE CATHETER**

To derive the model of behavior during active bending, we made the three following assumptions:

- When the active structure is completely bended (the bending angle is θmax), the SMA material is fully contracted and has reached a complete austenite state and thus presents an homogeneous Young modulus EA corresponding to this state,

- The bending of the structure follows the Euler-Bernouilli equation relating the local curvature ∂θ/∂L to the applied local bending moment M:

$$EI\frac{\partial\theta}{\partial L} = M \qquad \text{(A.1)}$$

  where E and I are respectively the global Young modulus and the quadratic moment of the bended structure,

- The central line of the external catheter and the central line of the sub-structure composed of the two metallic wires #1 and #2, remain very close to each other even during the bending phase. We call this common central line the central line of the structure,

- The compression forces applied to the structure during the bending phase have a negligible effect and thus the type of solicitation is close to pure bending moment. This implies, among others, that the central line of the structure remains the neutral (undeformed) line of the whole device during its bending.

Besides, in (30) we demonstrated that the distributed bending moment M applied along a bended structure by an SMA wire is constant and amounts to the product:

$$M = r.T \qquad \text{(A.2)}$$

where T stands for the tensile strength undergone by the contracted SMA wire and r is the distance between the SMA wire axis and the axis of the bended structure.

Moreover, the tensile strength T in the fully bended configuration is related to EA, τmax and SA the cross-sectional area of the SMA by the Hook's law of the SMA material in its fully austenitic state:

$$T = E_A S_A \tau = E_A \frac{\pi}{4} d^2 \tau \qquad \text{(A.3)}$$

By combining equations (3) and (A.1) to (A.3), and taking into account that

$$\frac{\partial \theta}{\partial L} = \frac{\theta_{max}}{L} = \frac{1}{R}$$

because the bending shape is circular as demonstrated in (30), we obtain the following expression for the radius of curvature of the bended structure:

$$R = \frac{1}{\tau_{lim}} \left( \frac{EI}{r E_A S_A} + \frac{rL}{L + L_{rig}} \right) \qquad \text{(A.4)}$$

As EI reflects in (A.4) the total rigidity of the bended structure, it can be expressed using the Young modulus Emet and quadratic moments Imet of the two metallic wires and the Young modulus Ec and quadratic moment Ic of the external catheter. Accounting for $I_{met} = \frac{\pi}{4} r_{met}^4$ and $I_c = \frac{\pi}{4}(r_{ext}^4 - r_{int}^4)$, this leads to:

$$EI = \frac{\pi}{4}(E_C(r_{ext}^4 - r_{int}^4) + 2.E_{met}.r_{met}^4) \ (A.5)$$

Finally, the radius of curvature of a bended section when its SMA wire is fully austenite amounts to:

$$R_{min} = \frac{1}{\tau_{lim}}\left(\frac{E_C(r_{ext}^4 - r_{int}^4) + 2.E_{met}.r_{met}^4}{r E_A d^2} + \frac{rL}{L + L_{rig}}\right) \quad (A.6)$$

And, combining (1) and (A.6), we find for the corresponding bending angle:

$$\theta_{max} = L.\tau_{lim}\left(\frac{E_C(r_{ext}^4 - r_{int}^4) + 2.E_{met}.r_{met}^4}{r E_A d^2} + \frac{rL}{L + L_{rig}}\right)^{-1} \quad (A.7)$$

**Figure Captions List**

Fig. 1          Extracorporeal robots: a- Magellan®, b- Niobe® systems

Fig. 2          SMA actuated catheter (principle) (21)

Fig. 3          Active catheter featured with three SMA actuators (21)

Fig. 4          Radius of curvature vs number of SMA actuators

Fig. 5          Manoeuver for entering a lateral artery a- general descrition, b-

                pointing toward the ostium, c- entering into the ostium, d- case of a

                large original artery

Fig. 6          Geometrical description of the device comprising of two active

                sections; a- general description; b- detail of an active section

Fig. 7          Relation between the angle and the different radii of curvature

Fig. 8          Cross-sectional view of the structure

Fig. 9          Axial anchoring of the SMA wire

Fig. 10         Radial anchoring of the SMA wire

Fig. 11         The active device and its external covering catheter

Fig. 12         Cross-sectional dimensions of the flexible part

**Table Caption List**

Table 1   Performances of small dimeter active catheters using SMA actuators

Table 2   Min and max diameters of the arteries of interest

Table 3   Min and max bifurcation angles for the region of interest

Table 4   Functional specifications

Table 5   Tested values of the geometric parameters

Table 6   Comparison between the initial functional specifications and the

characteristics of the best prototype